\documentclass[conference]{IEEEtran}
\usepackage{cite}
\usepackage{amsfonts,amssymb,amsthm,amsbsy,amsmath,paralist,bm,ifthen,color}
\usepackage{algorithm,algorithmic}
\usepackage{graphicx}
\usepackage[font=footnotesize]{caption}
\usepackage{xcolor}
\usepackage{relsize}
\usepackage[numbers,sort&compress,square]{natbib}
\usepackage{tabularx}
\usepackage{setspace}
\newtheorem{lemma}{Lemma}
\IEEEoverridecommandlockouts
\begin{document}
\captionsetup[figure]{labelfont={bf},name={Fig.},labelsep=period}
\title{Robust URLLC Packet Scheduling of OFDM Systems}
\author{Jing Cheng$^\dag$, Chao Shen$^\dag$, Shuqiang Xia$^\ddag$\\
{\small $^\dag$State Key Laboratory of Rail Traffic Control and Safety, Beijing Jiaotong University, Beijing, China}\\
{\small $^\ddag$Algorithm Department, Wireless Product R$\&$D Institute, ZTE Corporation, Shenzhen, China}\\
{\small Email: \{chengjing, chaoshen\}@bjtu.edu.cn, xia.shuqiang@zte.com.cn}
\thanks{The work was supported in part by the NSFC, China under Grant 61871027 and Grant U1834210, in part by the State Key Laboratory of Rail Traffic Control and Safety under Grant RCS2019ZZ002.}}
\maketitle

\begin{abstract}
In this paper, we consider the power minimization problem of joint physical resource block (PRB) assignment and transmit power allocation under specified delay and reliability requirements for ultra-reliable and low-latency communication (URLLC) in downlink cellular
orthogonal frequency-division multiple-access (OFDMA) system. To be more practical, only the imperfect channel state information (CSI) is assumed to be available at the base station (BS). The formulated problem is a combinatorial and mixed-integer nonconvex problem and is difficult to tackle. Through techniques of slack variables introduction, the first-order Taylor approximation and reweighted $\ell_1$-norm, we approximate it by a convex problem and the successive convex approximation (SCA) based iterative algorithm is proposed to yield sub-optimal solutions. Numerical results provide some insights into the impact of channel estimation error, user number, the allowable maximum delay and packet error probability on the required system sum power.
\end{abstract}

\section{Introduction}
 Ultra-reliable and low-latency communication (URLLC) is one of the areas of interest in B5G and 6G. It acts as an enabler for mission-critical applications such as factory automation, remote control, and intelligent transportation system (ITS). Such scenarios call for ultra-low latency (e.g. user-plane latency $1$ ms) and ultra-high reliability (e.g. $99.999\%$) \cite{8869705}, which imposes new challenges to the future cellular network. Although orthogonal frequency-division multiple-access (OFDMA) has been widely adopted as the air interface in wireless networks, enabling URLLC needs scalable numerology (new OFDMA based frame structure) rather than fixed numerology used in the LTE system \cite{8474959}. Based on the new frame structure of $5$G, the problem of packet scheduling and resource allocation to meet multiple URLLC users' quality-of-service (QoS) requirements (delay, reliability and transmission rate) arises.

 A number of resource allocation problems in the OFDMA system have been investigated in the literature to improve system performance. To exploit the space-time-frequency domain, the resource allocation problem has been considered in the multiple-input single-output (MISO) OFDMA and multiple-input multiple-output (MIMO) OFDMA system in \cite{5456049,7442570}, respectively. Given that the perfect channel state information (CSI) cannot be available at the transmitter side, authors in \cite{5961646,6786058} took into account the impact of imperfect CSI on the design of resource allocation scheme. In general, the resource allocation problem is jointly optimized with other aspects. For example, in \cite{8086180}, the prioritized link scheduling, subchannel assignment, and power allocation were jointly optimized to maximize the number of scheduled non-prioritized links and their sum rate under the minimum rate requirements of all links. Driven by the demand of URLLC, some researches investigated how to conduct flexible resource allocation design to meet such rigorous demands. The authors in \cite{8490699} considered the resource allocation problem both in the one-shot transmission scheme and the hybrid automatic repeat request (HARQ) scheme to maximize the admissible URLLC packets load. In \cite{8541123}, the globally optimal solutions of the joint uplink and downlink resource allocation problem for improving the energy efficiency
 can be obtained for a URLLC system. However, the above two URLLC-relevant works assumed that the channel gain is same even for different subchannels. The recent work \cite{8756746} has considered that factor and formulated a resource allocation problem for multi-user downlink URLLC-OFDMA system to maximize the weighted sum throughput with constraints on the transmission bit number, delay and reliability. Unlike the well studied weighted sum throughput maximization problem with perfect CSI in \cite{8756746}, in this paper, we focus on
 the sum power minimization problem with imperfect CSI and specified URLLC demands since the assumption of imperfect CSI is more practical especially in URLLC systems. In view of this, the worst-case robust packet scheduling and resource allocation is of significant importance in URLLC applications.

Therefore, in this paper, we consider the joint optimization problem of physical resource block (PRB) assignment and power allocation to minimize the total transmit power of the base station (BS) in the downlink cellular URLLC-OFDMA system, while guarantee the specified QoS requirements of stringent delay, ultra-high reliability and minimum number of required transmission bits for each URLLC user. The main contributions of this paper are summarized as
follows.
\begin{itemize}
  \item We propose a worst-case robust scheduling and transmission scheme to guarantee the specified QoS requirements
  in a multi-user URLLC-OFDMA system.
  \item The formulated mixed-integer non-convex problem is challenging to resolve. By utilizing the structure of the problem, we reformulate it into a new problem defined in a larger feasible region and then the original solutions can be recovered.
  \item By virtue of introduction of slack variables, the first-order Taylor approximation and reweighted $\ell_1$ approximation, we transform it into a convex problem. Due to the high complexity of getting the globally optimal solution, a successive convex approximation (SCA) based iterative algorithm is proposed to efficiently obtain at least a stationary point of the problem.
\end{itemize}
\section{System and Channel Models}
In this section, we present the system and channel models.
\subsection{System Model}
Consider a single-cell downlink multi-user transmission scenario, where a single-antenna BS serves $K$ single-antenna URLLC users indexed by $k\in \mathcal{K}\triangleq\{1,\cdots,K\}$. Based on the $5$G NR frame structure \cite{3gppTR21915}, we consider the PRB as a fundamental scheduling resource unit, where each resource block consists of $12$ consecutive subcarriers and one slot assumed to span one OFDM symbol duration for simplicity. There are totally $M\times N$ PRBs\footnote{A typical subcarrier bandwidth of  $15$ kHz corresponds to $66~\mu s$ for each OFDM symbol in $5$G NR, therefore, $N$ should be less than $7$ to meet the URLLC delay requirement of $1$ ms.} for scheduling in the time-frequency plane, where $M$ and $N$ denoted the number of frequency bins and slots, respectively. As for the $k$th user, the BS has to send
it a URLLC packet of $B_k$ information bits within $D_k$ time slots and with maximum reliability $1-\varepsilon_k$. We assume the requirements $\{B_k,D_k,\varepsilon_k\}_{k=1}^K$ of all users are known at the BS.
In order to indicate on which PRBs a URLLC packet is transmitted, we define a binary assignment indicator
$I_{mnk}\in\{0,1\}$ where $m\in \mathcal{M}\triangleq\{1,\cdots,M\},~n\in \mathcal{N}\triangleq\{1,\cdots,N\}$. If the frequency bin $m$ in time slot $n$ is allocated to user $k$, we have $I_{mnk}=1$, otherwise $I_{mnk}=0$. Furthermore, we assume that each PRB is allotted to at most one user to avoid inter-user interference.
\subsection{Channel Model}
We assume the channel is quasi-static block fading, which means that the channel remains constant within each coherent block, and varies independently across blocks. However, due to the practical limitations, e.g., CSI estimation error, quantization error,  CSI feedback delay, it is impossible to obtain perfect CSI at the transmitter side. This is true specifically for the mission critical scenarios where the time for channel training is highly restricted for data transmission.

In this paper, we adopt the bounded CSI error model to characterize the channel uncertainty \cite{6156468}. Specifically, the actual channel between the BS and the $k$th user on the frequency bin $m$ and time slot $n$ can be modeled as
\begin{equation}
  h_{mnk}=\hat{h}_{mnk}+e_{mnk},~|e_{mnk}|\leq \delta_{mnk},
\end{equation}
where $\hat{h}_{mnk}$ denotes the channel estimate, $e_{mnk}$ is the channel estimation error, and $\delta_{mnk}$ defines the bound of the uncertainty region.

Under this bounded channel error model, the received signal-to-noise ratio (SNR) of the $k$th user in the worst case can be expressed as
\begin{equation}
  \rho_{mnk}=\min_{|e_{mnk}|\leq \delta_{mnk}}\frac{\alpha_k \left|\hat{h}_{mnk}+e_{mnk}\right|^2 P_{mnk}}{\sigma^2},
\end{equation}
where $\alpha_k$ is the large-scale channel gain at user $k$ that depends on the path loss and shadowing, $P_{mnk}$ is the transmit power assigned to user $k$ on frequency bin $m$ and time slot $n$, and $\sigma^2$ is the power of additive white Gaussian noise (AWGN). It can be easily obtained that $e_{mnk}^{*}=-\frac{\hat{h}_{mnk}}{\left|\hat{h}_{mnk}\right|}\delta_{mnk}$, thus the received SNR of user $k$ can be rewritten as
\begin{equation}
  \rho_{mnk}=c_{mnk}P_{mnk},
\end{equation}
where $c_{mnk}\triangleq \frac{\alpha_k\left(\left|\hat{h}_{mnk}\right|-\delta_{mnk}\right)^2}{\sigma^2}$ for all $m,n,k$.
\section{Robust Scheduling Problem Formulation}
\subsection{Achievable Rate Characterization}
Typically, small packets are required to transmit in URLLC systems with extremely stringent latency constraints, thereby rendering the traditional Shannon's capacity unapplicable to characterize the maximum achievable rate \cite{8624559,8345745,8909370}. Y. Polyanskiy etc. proposed a formula for interference-free flat-fading channel which characterizes the relationship between the latency, decoding error probability and achievable rate in the finite blocklength regime \cite{5452208}. The achievable rate with finite blocklength (FBL) code has been  extended to different channels, including the fading channel \cite{6802432,8756746,8070468}.

Here we consider the joint channel coding scheme where the data packet of each user is encoded over all scheduled PRBs. Thus, the maximum number of received data bits $R_k$ for user $k$ can be approximately characterized by
\begin{equation}
  R_k \!= \sum_{m=1}^M \sum_{n=1}^N I_{mnk}\log_2(1+\rho_{mnk})\!-\!\sqrt{x_k}\frac{Q^{-1}(\varepsilon_k)}{\ln 2},
\end{equation}
where
\begin{equation}
  x_k=\sum_{m=1}^M \sum_{n=1}^N I_{mnk}V_{mnk},
\end{equation}
$Q^{-1}(\cdot)$ is the inverse of $Q(x)=\int_{x}^{\infty} \frac{1}{\sqrt{2 \pi}} e^{-t^{2}/2}dt$, and $V_{mnk}=1-\frac{1}{\left(1+\rho_{mnk}\right)^2}$ is the channel dispersion.

\begin{figure}
	\centering
	\includegraphics[width=.9\linewidth]{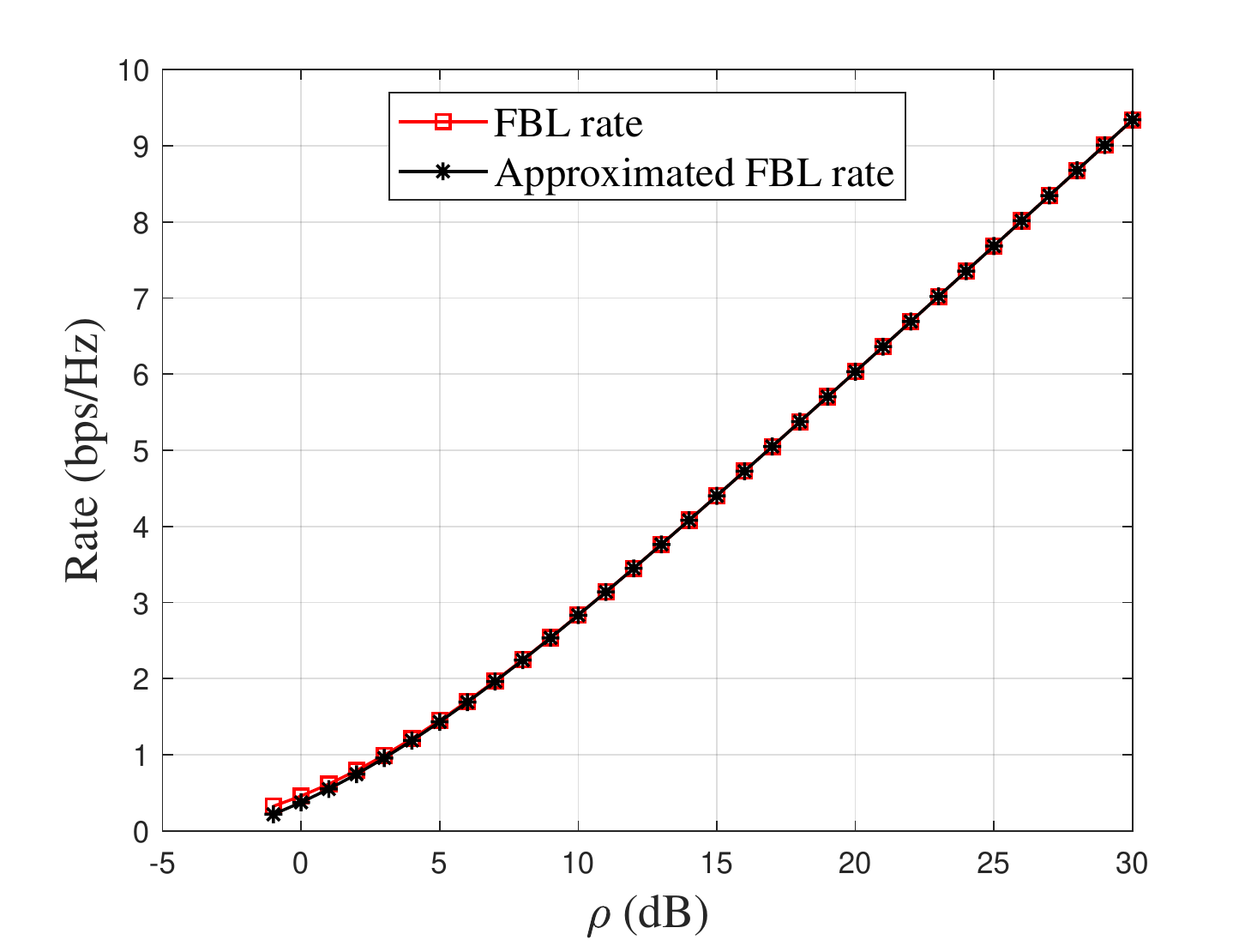}
	\caption{Contrast of FBL rate and approximated FBL rate.}\vspace{-5mm}
	\label{approximationFBLCapacity}
\end{figure}

Note that the approximation $V_{mnk}\approx 1$ is adopted in this paper since this approximation is accurate enough when the received SNR is higher than $3$ dB, and it is almost true for URLLC scenarios. Here, we plot a picture to verify the accuracy of this approximation for $120$ channel uses and decoding error probability $\varepsilon=10^{-6}$, as shown in Fig.\ref{approximationFBLCapacity}.

\subsection{Optimization Problem Formulation}
In this paper, we are interested in (1) how much power the BS needs to support all users in the system to complete transmission tasks with requirements $\{B_k,D_k,\varepsilon_k\}_{k=1}^K$; (2) how to perform resource scheduling to minimize the power consumption of the BS under the imperfect CSI, thereby maximizing energy efficiency. Motivated by these interests, the power minimization problem of robust scheduling and resource allocation for URLLC packets can be formulated as the following optimization programming
\begin{subequations}\label{P1}
\begin{align}
\min_{\{I_{mnk}, P_{mnk}\}}~&P_{\text{tot}}\triangleq\sum_{k=1}^K\sum_{m=1}^M\sum_{n=1}^N I_{mnk}P_{mnk}\label{P1a}\\
\text{s.t.}\quad\quad~ &R_k\geq B_k,~\forall k,\label{P1b}\\
&I_{mnk}\in \{0,1\},~\forall m,n,k,\label{P1c}\\
&\sum_{k=1}^K I_{mnk} \leq 1,~\forall m,n,\label{P1d}\\
&I_{mnk}=0,~\forall n>D_k,~\forall m,k,\label{P1e}\\
&0\leq P_{mnk}\leq I_{mnk}P_{\max},~\forall m,n,k.\label{P1f}
\end{align}
\end{subequations}
 The constraint in \eqref{P1b} guarantees the target payload demand of $B_k$ bits for user $k$. Constraint \eqref{P1c} and \eqref{P1d} require that each PRB is allocated to at most one user.
 The delay requirement that the packet $k$ has to be successfully transmitted within $D_k$ time slots is reflected in \eqref{P1e}. The constraint \eqref{P1f} means the nonnegative power constraint and guarantees that $P_{mnk}=0$ if $I_{mnk}=0$, which is in accordance with engineering practice.

Problem \eqref{P1} is a combinatorial and mixed-integer nonconvex problem. Its main challenges are due to the binary indicator and the strict nonconvexity of the constraint \eqref{P1b}. The problem \eqref{P1} is NP-hard in general \cite{7463506,8269136} and the globally optimal solution by exhaustive search incurs high computational complexity. Hence, it is necessary to develop an algorithm to approximately solve the problem \eqref{P1} such that the URLLC packet scheduling can be efficiently performed.
\section{SCA-based Robust Transmission Scheme}
In this section, we first reformulate the optimization problem \eqref{P1} as a new one which can always yield an optimal solution to \eqref{P1} if it is feasible. And then we introduce some slack variables and the reformulated problem can thus be approximately resolved by applying the SCA and reweigthed $\ell_1$-norm techniques.

By capturing the existence of the product term $I_{mnk}P_{mnk}$, we exert some transformations on the constraints \eqref{P1b} and \eqref{P1f} as follows
\begin{align}
  &\sum_{m=1}^{M} \sum_{n=1}^{N} I_{mnk} \log _{2}\left(1+\frac{c_{mnk}I_{mnk} P_{mnk}}{I_{mnk}}\right)\notag\\
  &~~~~~~~~~~~~~~~~~~~~~~~-\sqrt{x_k} \frac{Q^{-1}\left(\varepsilon_{k}\right)}{\ln 2}\geq B_k,~\forall k,\label{TC1}
\end{align}
\begin{equation}
  0 \leq I_{mnk}P_{mnk} \leq I_{mnk} P_{\max }, \forall m, n, k.\label{TC2}
\end{equation}
Then the original optimization problem \eqref{P1} can be rewritten as
\begin{subequations}\label{P2}
\begin{align}
\min_{\{I_{mnk}, P_{mnk}\}}~&P_{\text{tot}}\\
\text{s.t.}\quad\quad~ &\eqref{P1c}-\eqref{P1e},\eqref{TC1},\eqref{TC2}.
\end{align}
\end{subequations}
First, we have the following lemma for the problem \eqref{P1}.
\begin{lemma}\label{lemma1}
  The optimal solution to \eqref{P1} is optimal to the problem \eqref{P2}, and can be restored from the optimal solution to \eqref{P2} if it is feasible.
\end{lemma}
\proof Note that the feasible set of the problem \eqref{P2} is larger than the original problem \eqref{P1} due to the difference between the constraint \eqref{P1f} and \eqref{TC2}. Specifically, when $I_{mnk}=0$, \eqref{P1f} means $P_{mnk}=0$ whereas $P_{mnk}$ can take any value based on \eqref{TC2}. When it comes to the case $I_{mnk}=1$, constraint \eqref{P1f} and \eqref{TC2} are equivalent. It is necessary to remark that based on the optimal solution to the problem \eqref{P2} (if feasible), the optimal solution to the problem \eqref{P1} is recovered by letting $P_{mnk}=0$, if $I_{mnk}=0$. Hence, the problem \eqref{P2} is a positive alternative to the problem \eqref{P1}.   \hfill $\blacksquare$

With the Lemma \ref{lemma1}, we can then introduce a slack variable $p_{mnk}$ to resolve the challenge caused by the coupling between $I_{mnk}$ and $P_{mnk}$. By letting $p_{mnk}$ equal to $I_{mnk}P_{mnk}$, we reformulate the problem \eqref{P2} as
\begin{subequations}\label{P3}
\begin{align}
\min_{\{I_{mnk},p_{mnk}\}}~&p_{\text{tot}}\triangleq\sum_{k=1}^K\sum_{m=1}^M\sum_{n=1}^N p_{mnk}\label{P3a}\\
\text{s.t.}\quad\quad~ &\sum_{m=1}^{M} \sum_{n=1}^{N} I_{mnk} \log _{2}\left(1+\frac{c_{mnk}p_{mnk}}{I_{mnk}}\right)\notag\\
  &~~~~~~~~~~~~~~~~~~~~-\sqrt{x_k} \frac{Q^{-1}\left(\varepsilon_{k}\right)}{\ln 2}\geq B_k,~\forall k,\label{P3b}\\
&0 \leq p_{mnk} \leq I_{mnk} P_{\max }, \forall m, n, k,\label{P3d}\\
&\eqref{P1c}-\eqref{P1e}.\label{P3c}
\end{align}
\end{subequations}

Note that the constraint \eqref{P3b} is still nonconvex. We resort to the first-order Taylor expansion to approximate it. For constraint \eqref{P3b}, the first term in the left-hand side is concave since perspective function preserves convexity. Then, by the first-order Taylor approximation of the concave function $\sqrt{x_k}$, we further obtain the following locally tight upper bound $\frac{x_k+x_k^{(i)}}{2\sqrt{x_k^{(i)}}}$, where $x_k^{(i)}$ is the value of variable $x_k$ in the $i$th iteration. Based on the above approximation, the constraint \eqref{P3b} can be converted into a convex constraint as

\begin{align}
&\sum_{m=1}^{M}\sum_{n=1}^{N} I_{mnk}\log _{2}\left(1+\frac{c_{mnk}p_{mnk}}{I_{mnk}}\right)\notag\\
&~~~~~~~~~~~~~~~~~~~~~~~-\frac{x_k+x_k^{(i)}}{2\sqrt{x_k^{(i)}}}\frac{Q^{-1}(\varepsilon_k)}{\ln 2}\geq  B_k,\forall k.\label{Bit}
\end{align}

It's worth noting that it's the binary nature of the scheduling variables $I_{mnk},\forall m,n,k$ that makes the problem intractable. Therefore, we relax them into continuous variables between $0$ and $1$. Considering the requirement that each PRB is allocated to at most one user, we have to add the constraint $\|\mathbf{I}_{mn}\|_0\le 1,\forall m,n$ to guarantee this sparsity requirement, where $\mathbf{I}_{mn}\in \mathbb{R}_{+}^K$ is defined as $\mathbf{I}_{mn}=[I_{mn1},\cdots,I_{mnK}]^T$. Then, the problem in each SCA iteration can be formulated as
\begin{subequations}\label{P4}
\begin{align}
\min_{\{I_{mnk},p_{mnk}\}}~&p_{\text{tot}}\\
\text{s.t.}\quad\quad~ &0\leq I_{mnk}\leq 1,~\forall m,n,k,\label{P4b}\\
&\|\mathbf{I}_{mn}\|_0\le 1,~\forall m,n,\label{P4c}\\
&\eqref{P1d},\eqref{P1e},\eqref{P3d},\eqref{Bit}.
\end{align}
\end{subequations}

Eventually, given that the problem \eqref{P4} is nonconvex owing to $\ell_{0}$-norm in the constraint \eqref{P4c}, we exploit the reweighted $\ell_1$-norm to approximate it \cite{Cand2007Enhancing}, which can enhance the sparsity of solutions and improve the performance of $I_{mnk}$ recovery for all $m,n,k$. For any $m,n$, define a weight matrix $\mathbf{W}_{mn}\in \mathbb{R}_{+}^{K\times K}$, which is a diagonal matrix with $\{W_{mn1},\cdots,W_{mnK}\}$ on the diagonal and zeros elsewhere. Based on the method in \cite{Cand2007Enhancing}, we construct the weights $W_{mnk}=\frac{1}{I_{mnk}+\xi}$ for each $k=1,\cdots,K$, where $\xi>0$ is a relatively small value. Note that such construction approach enforces the nonzero elements of $\mathbf{I}_{mn},\forall m,n$ level off to $1$. This reweighted $\ell_1$-norm approximation produces a convex constraint set as
\begin{equation}
  \|\mathbf{W}_{mn}\mathbf{I}_{mn}\|_1 \leq 1,~\forall m,n. \label{RLC}
\end{equation}
Thus, the optimization problem \eqref{P4} can be approximated into a convex one, given by
\begin{subequations}\label{P5}
\begin{align}
\min_{\{I_{mnk},p_{mnk}\}}~&p_{\text{tot}}\\
\text{s.t.}\quad\quad~ &\eqref{P1d},\eqref{P1e},\eqref{P3d},\eqref{Bit},\eqref{P4b},\eqref{RLC}, \end{align}
\end{subequations}
which can be efficiently solved with the off-the-shelf convex solver like CVX. As a result, the suboptimal solution to the optimization problem \eqref{P1} can be obtained by iteratively solving \eqref{P5} until convergence.

According to the preceding analysis, now we propose an iterative algorithm to solve the optimization problem \eqref{P1} by applying SCA \cite{6365845}. The SCA-based algorithm is summarized in Algorithm \ref{Alg1}. It can be proved by \cite{Razaviyayn2012A} that the proposed algorithm can converge to at least a stationary solution if it is feasible.
\begin{algorithm}
		\caption{\small \textbf{:} SCA-based Algorithm for Solving Problem \eqref{P1}}
		\label{Alg1}
\begin{spacing}{1.2}
		\begin{algorithmic}[1]
			\STATE ~$i=0$,~$\Delta=1$, tolerance $\epsilon=10^{-6}$, $p_{\text{tot}}^{(i)}=0$, $\xi=0.01$.
			\STATE \textbf{Initialize:}~$W_{mnk}^{(i)} = 1$ and a feasible set  $I_{mnk}^{(i)},~\forall m,n,k$.
			\STATE \textbf{Calculate} $x_k^{(i)} = \sum_{m=1}^{M} \sum_{n=1}^{N} I_{m n k}^{(i)}$, for all $k$.
			\WHILE{$\Delta \geq \epsilon$}
			\STATE Obtain $p_{\text{tot}}^{(i+1)}$ and $\{I_{mnk}^{(i+1)},p_{mnk}^{(i+1)}\}$ by solving \eqref{P5}.			
			\STATE Update $x_k^{(i+1)}:=\sum_{m=1}^M\sum_{n=1}^N I_{mnk}^{(i+1)}$.
			\STATE Update $W_{mnk}^{(i+1)}:=\frac{1}{I_{mnk}^{(i+1)}+\xi}$.
			\STATE Update $\Delta:=\left|p_{\text{tot}}^{(i+1)}-p_{\text{tot}}^{(i)}\right|$,
			\STATE $i:=i+1$.
			\ENDWHILE
			\STATE \textbf{Output} $P_{mnk}^{*}=p_{mnk}^{(i)}$ if $I_{mnk}^{(i)}=1$, and 0 otherwise; $I_{mnk}^{*}=I_{mnk}^{(i)}$.
		\end{algorithmic}
\end{spacing}
\end{algorithm}
\section{Numerical Results}
In this section, we show some simulation results to evaluate the performance of the proposed robust scheduling and resource allocation scheme.

In the simulation, the cell radius is $200$ m. We consider the worst case that all URLLC users are located at the edge of the cell. We denote by $\hat{h}_{mnk}\sim\mathcal{CN}(0,1),~\forall m,n,k$ the Rayleigh fading component. Moreover, the path loss component is modeled as $35.3+37.6\log_{10}(d_k)$ in dB \cite{8541123}, where $d_k$ in meter denotes the distance from the BS to the user $k$. The number of frequency bins is $64$ and the bandwidth of each PRB is $180$ kHz. The power spectral density of the AWGN at the users is assumed to be $-169$ dBm/Hz. The parameter $\xi$ for constructing the weight matrix is $0.01$. We generate $100$ channel realizations and take their average as each simulation result.

\begin{figure}
	\centering
	\includegraphics[width=.9\linewidth]{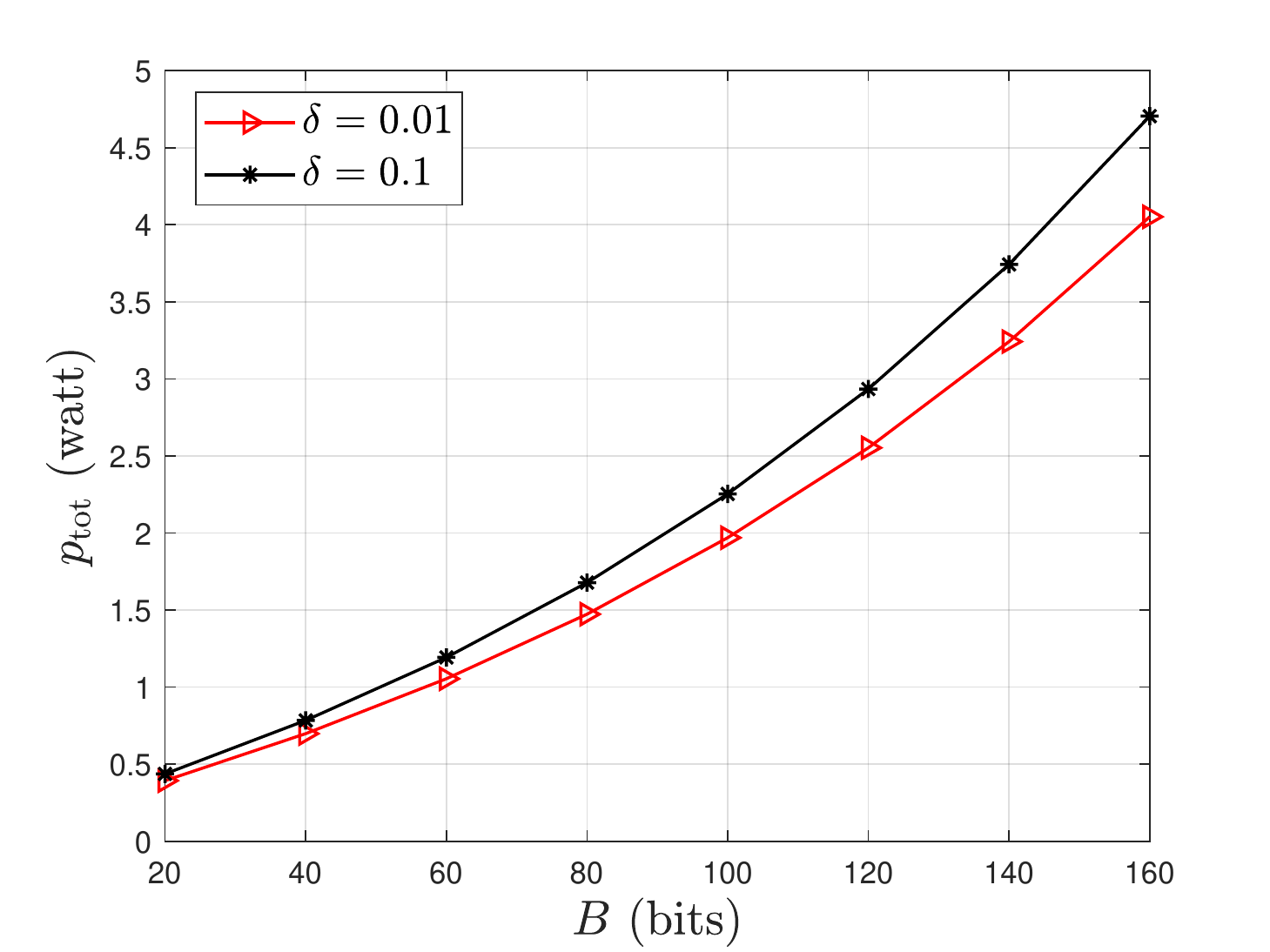}
	\caption{Sum Power $p_\text{tot}$ versus transmission bits $B$.}\vspace{-5mm}
	\label{sumPower_vs_Bit}
\end{figure}

Fig.\ref{sumPower_vs_Bit} shows the minimum sum power $p_\text{tot}$ required to transmit target payload $B_k=B$ bits $\forall k$ with different channel estimation error $\delta$ for $P_{\max}=23$ dBm. There are $4$ URLLC users and the number of time slots $N$ is $6$. The delay of these four users is set as $D_1=3,D_2=D_3=4,D_4=6$, respectively. We set the maximum packet error probability $\varepsilon_k=10^{-6},\forall k$. It is observed that $p_\text{tot}$ grows almost exponentially with the increase of the minimum transmission bits $B$. Note that the larger the channel estimation error $\delta$ is, the greater the sum power required to transmit the same number of bits is, and the impact of the channel estimation error is comparatively greater when the target payload is large (e.g. $160$ bits).

\begin{figure}
	\centering
	\includegraphics[width=.9\linewidth]{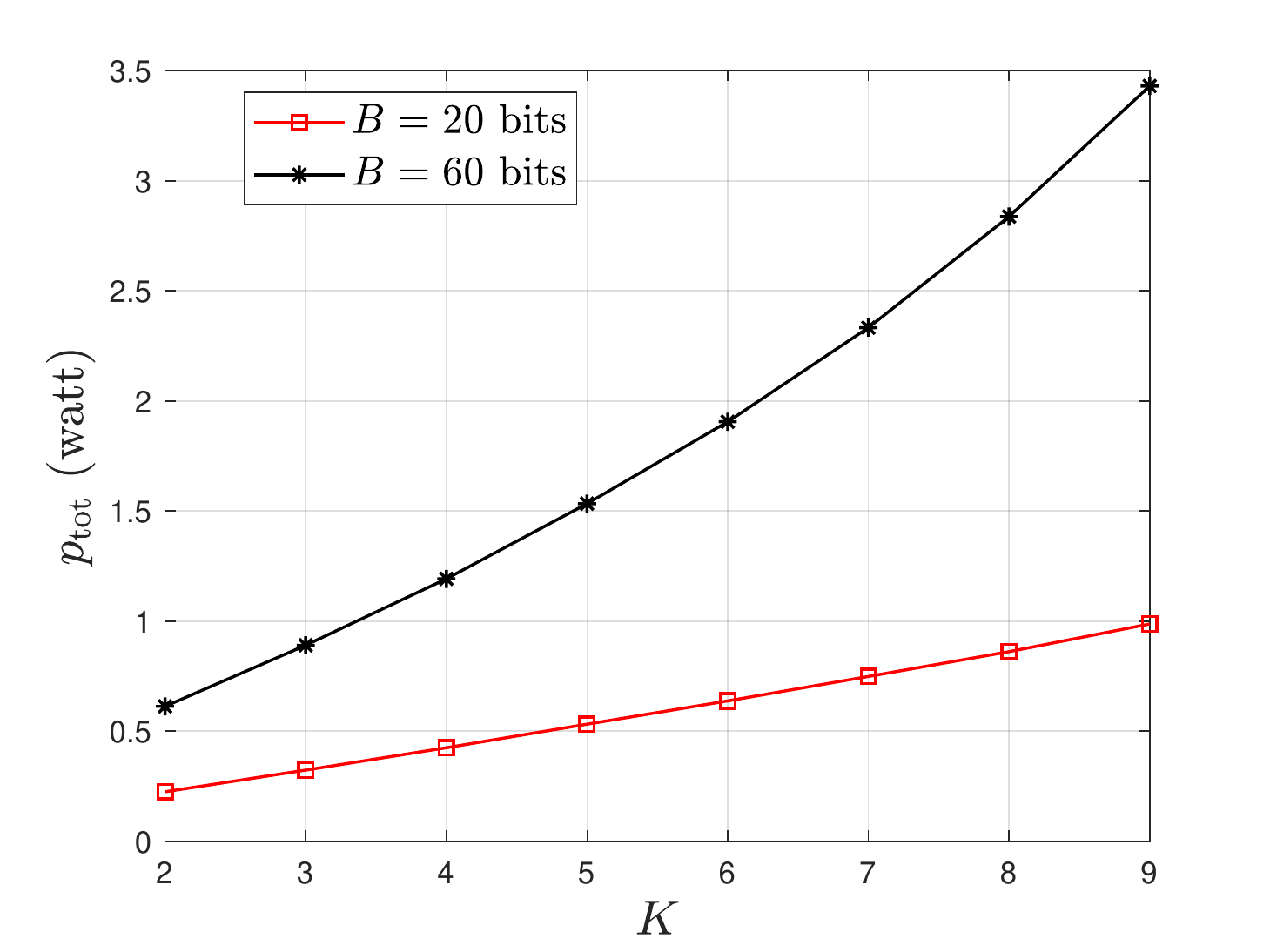}
	\caption{Sum Power $p_\text{tot}$ versus number of users $K$.}\vspace{-5mm}
	\label{sumPower_vs_NumberofUser}
\end{figure}

In Fig.\ref{sumPower_vs_NumberofUser}, we plot the minimum sum power $p_\text{tot}$ required by all users in the system to transmit the target number of bits $B_k=B,~\forall k$ within a given delay requirement $D_1=2,D_k=4,~\forall k\neq 1$ for $P_{\max}=38$ dBm, $\delta=0.01$ and $\varepsilon_k=10^{-6},\forall k$. The number of time slots $N$ is $4$. Obviously, when $K$ is given, the more $p_\text{tot}$ is required as the payload to be transmitted increases. As can be seen from  Fig.\ref{sumPower_vs_NumberofUser}, when the target transmission bits is small (e.g. $20$ bits), the sum power $p_\text{tot}$ grows linearly with the increase of the number of URLLC users $K$ in the system. In contrast, when the required transmission bits is considerable (e.g. $60$ bits), $p_\text{tot}$ grows exponentially with the increase of $K$. This result is in line with the fact that when the number of time-frequency resource blocks is fixed, with the increase of $K$, the allocation of larger power renders the satisfaction of minimum transmission payload requirements possible.

\begin{figure}
	\centering
	\includegraphics[width=.9\linewidth]{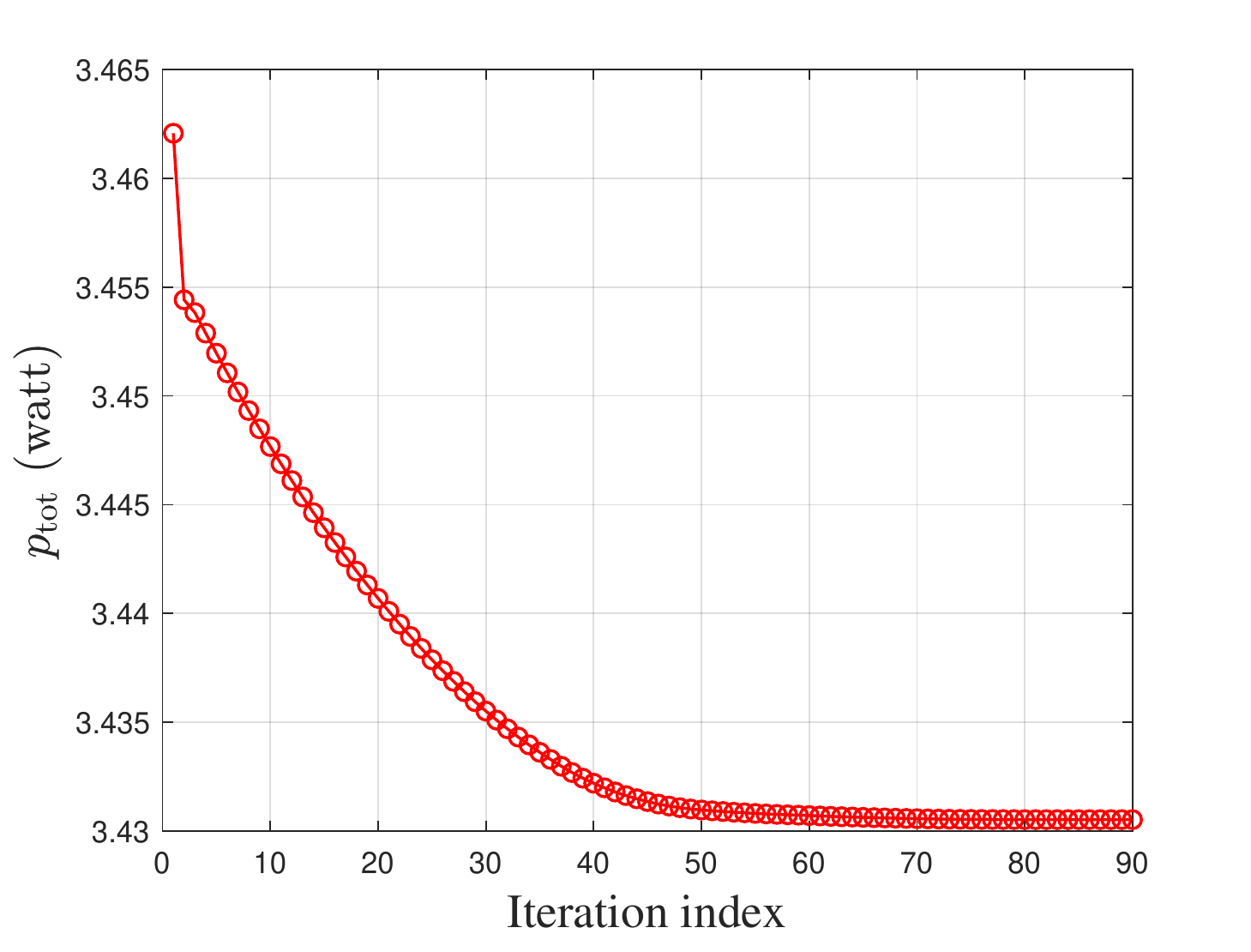}
	\caption{Convergence performance for $N=64,~K=9$.}\vspace{-5mm}
	\label{Convergence}
\end{figure}

In Fig.\ref{Convergence}, we evaluate the convergence performance of the proposed iterative algorithm when the number of frequency bins and users is large (e.g. $N=64,~K=9$). In this case, other parameters are set as $B_k=60 \text{bits},~\forall k,~P_{\max}=38$ dBm, $\delta=0.01,~\varepsilon_k=10^{-6},\forall k,~N=4$. Note that as the increase of the number of frequency bins and users, the proposed algorithm needs considerably more iterations to converge since additional users lead to additional search dimensions in the expanded time-frequency plane. As for the case of $N=64,~K=9$, this algorithm can reach convergence in 60 iterations on average.

\begin{figure}
	\centering
	\includegraphics[width=.9\linewidth]{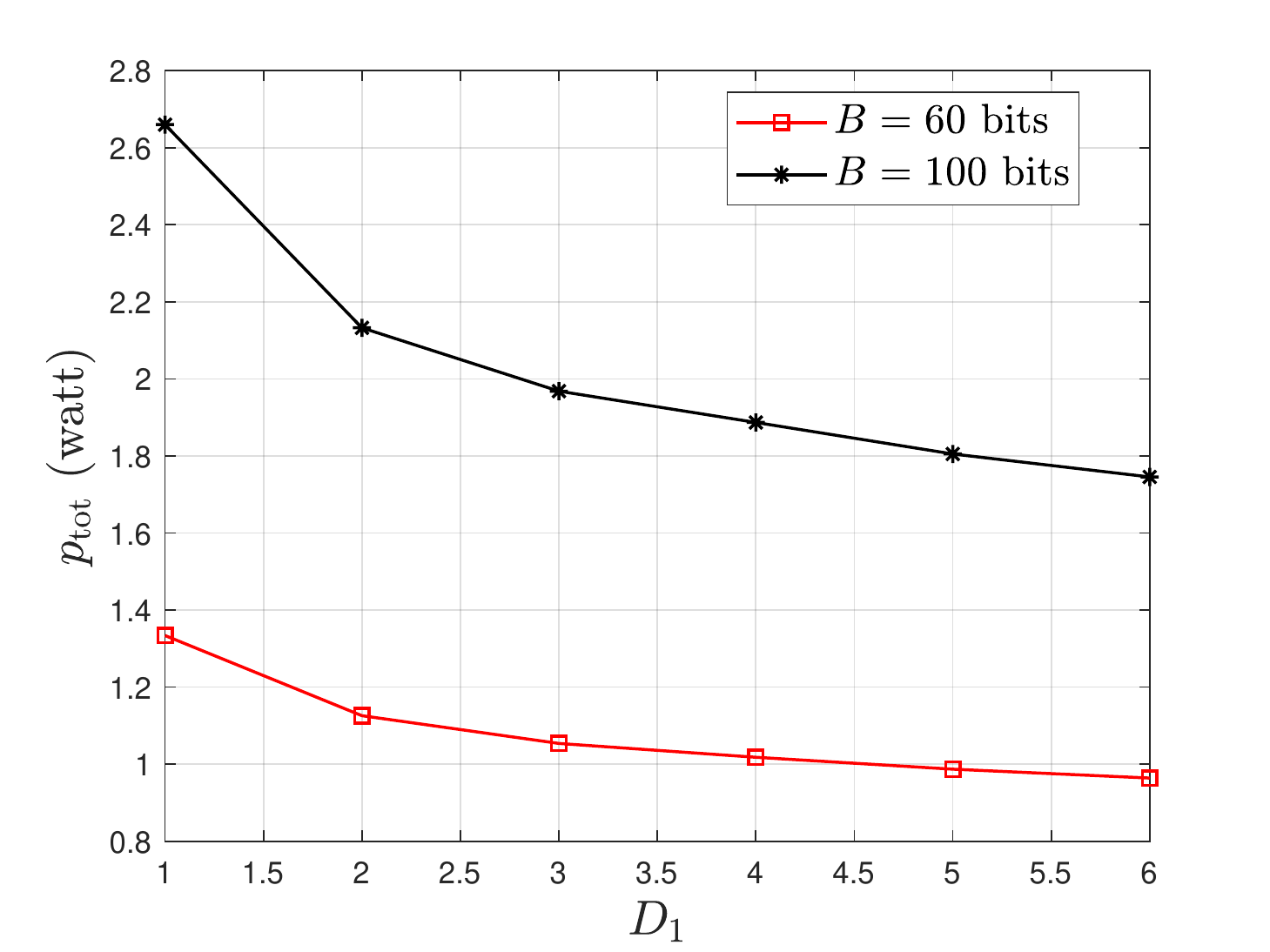}
	\caption{Sum Power $p_\text{tot}$ versus $D_1$.}\vspace{-5mm}
	\label{sumPower_vs_delay}
\end{figure}

Fig.\ref{sumPower_vs_delay} illustrates the influence of user's delay on the $p_\text{tot}$ required to complete the target transmission task $B_k=B,~\forall k$ for $P_{\max}=23$ dBm, $\delta=0.01,~\varepsilon_k=10^{-6},\forall k$. There are $4$ users and the number of time slots $N$ is $6$. In order to facilitate the observation of the effect of delay on the required power, we change the maximum allowable delay of the first user from $1$ to $6$ (e.g. $D_1\in[1,6]$), while the delay requirements of other users remain unchanged (e.g. $D_2=D_3=4,D_4=6$). The user's delay requirements have a greater impact on the power, which is more significant when the number of transmission bits is large. As can be seen from the Fig.\ref{sumPower_vs_delay}, if $D_1$ is relaxed from $1$ to $6$, $p_\text{tot}$ is reduced by about $0.4$ watt for $B=60$ bits, while $p_\text{tot}$ is sharply reduced by about $0.9$ watt for $B=100$ bits.

\begin{figure}
	\centering
	\includegraphics[width=.9\linewidth]{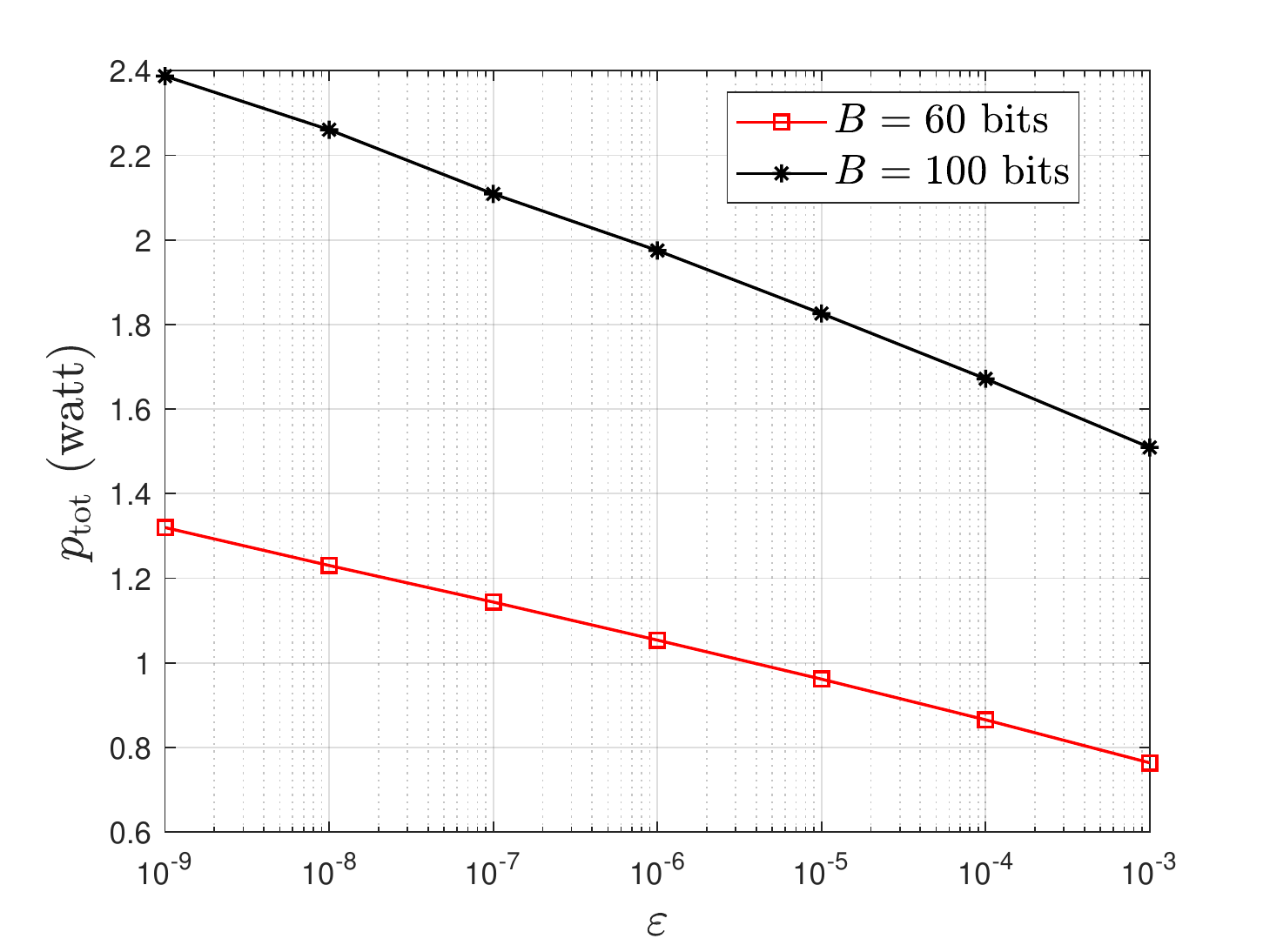}
	\caption{Sum Power $p_\text{tot}$ versus packet error probability $\varepsilon$.}\vspace{-5mm}
	\label{sumPower_vs_reliability}
\end{figure}

Fig.\ref{sumPower_vs_reliability} presents the minimum sum power $p_\text{tot}$ required to complete the transmission of  $B_k=B$ bits with a given maximum packet error probability $\varepsilon_k=\varepsilon,~\forall k$ for $P_{\max}=32$ dBm, $\delta=0.01$. There are $4$ users and the number of time slots $N$ is $6$. The delay of these four users is set as $D_1=3,D_2=D_3=4,D_4=6$, respectively. It is observed that the required sum power $p_\text{tot}$ decreases significantly as the allowable maximum packet error probability $\varepsilon$ increases, which is in accordance with the capacity formula in the FBL regime. Also, when $\varepsilon$ is given, the more $p_\text{tot}$ is required as the number of data bits to be transmitted increases.
 \section{Conclusion}
In this paper, we formulated the sum power minimization problem which jointly optimizes the allocation of PRBs and power allocation with imperfect CSI, specified QoS requirements for URLLC users. By using the structure of the original formulated problem, we find a positive alternative to it and the solutions to the original one can be recovered. By introducing slack variables and applying the first-order Taylor approximation and reweighted $\ell_1$-norm technique, we convert the nonconvex combinatorial and mixed-integer problem into a convex one. Then a low complexity SCA-based iterative algorithm is proposed. Numerical results illustrate that the channel estimation error and delay of users have a significant impact on the power required by the system especially when the transmission payload is large. Besides, when the time-frequency resource is fixed, more power needs to be allocated to meet the minimum transmission payload requirements as the number of users increases. Even when the number of frequency bins and users is large, the proposed SCA-based iterative algorithm can reach fast convergence.


\begin{thebibliography}{10}
\providecommand{\url}[1]{#1}
\csname url@samestyle\endcsname
\providecommand{\newblock}{\relax}
\providecommand{\bibinfo}[2]{#2}
\providecommand{\BIBentrySTDinterwordspacing}{\spaceskip=0pt\relax}
\providecommand{\BIBentryALTinterwordstretchfactor}{4}
\providecommand{\BIBentryALTinterwordspacing}{\spaceskip=\fontdimen2\font plus
\BIBentryALTinterwordstretchfactor\fontdimen3\font minus
  \fontdimen4\font\relax}
\providecommand{\BIBforeignlanguage}[2]{{%
\expandafter\ifx\csname l@#1\endcsname\relax
\typeout{** WARNING: IEEEtran.bst: No hyphenation pattern has been}%
\typeout{** loaded for the language `#1'. Using the pattern for}%
\typeout{** the default language instead.}%
\else
\language=\csname l@#1\endcsname
\fi
#2}}
\providecommand{\BIBdecl}{\relax}
\BIBdecl

\bibitem{8869705}
W.~{Saad}, M.~{Bennis}, and M.~{Chen}, ``A vision of 6{G} wireless systems:
  {A}pplications, trends, technologies, and open research problems,''
  \emph{IEEE Netw.}, pp. 1--9, 2019.

\bibitem{8474959}
K.~S. {Kim}, D.~K. {Kim}, C.~{Chae}, S.~{Choi}, Y.~{Ko}, J.~{Kim}, Y.~{Lim},
  M.~{Yang}, S.~{Kim}, B.~{Lim}, K.~{Lee}, and K.~L. {Ryu}, ``Ultrareliable and
  low-latency communication techniques for tactile internet services,''
  \emph{Proc. IEEE}, vol. 107, no.~2, pp. 376--393, Feb 2019.

\bibitem{5456049}
V.~D. {Papoutsis}, I.~G. {Fraimis}, and S.~A. {Kotsopoulos}, ``User selection
  and resource allocation algorithm with fairness in {MISO-OFDMA},'' \emph{IEEE
  Commun. Lett.}, vol.~14, no.~5, pp. 411--413, May 2010.

\bibitem{7442570}
G.~{Femenias} and F.~{Riera-Palou}, ``Scheduling and resource allocation in
  downlink multiuser {MIMO-OFDMA} systems,'' \emph{IEEE Trans. Commun.},
  vol.~64, no.~5, pp. 2019--2034, May 2016.

\bibitem{5961646}
R.~{Aggarwal}, M.~{Assaad}, C.~E. {Koksal}, and P.~{Schniter}, ``Joint
  scheduling and resource allocation in the {OFDMA} downlink: Utility
  maximization under imperfect channel-state information,'' \emph{IEEE Trans.
  Signal Process.}, vol.~59, no.~11, pp. 5589--5604, Nov 2011.

\bibitem{6786058}
Z.~{Chang}, T.~{Ristaniemi}, and Z.~{Niu}, ``Radio resource allocation for
  collaborative {OFDMA} relay networks with imperfect channel state
  information,'' \emph{IEEE Trans. Wireless Commun.}, vol.~13, no.~5, pp.
  2824--2835, May 2014.

\bibitem{8086180}
T.~D. {Hoang} and L.~{Bao Le}, ``Joint prioritized scheduling and resource
  allocation for {OFDMA}-based wireless networks,'' \emph{IEEE Trans. Wireless
  Commun.}, vol.~17, no.~1, pp. 310--323, Jan 2018.

\bibitem{8490699}
A.~{Anand} and G.~{de Veciana}, ``Resource allocation and {HARQ} optimization
  for {URLLC} traffic in 5{G} wireless networks,'' \emph{IEEE J. Sel. Areas
  Commun.}, vol.~36, no.~11, pp. 2411--2421, Nov 2018.

\bibitem{8541123}
C.~{Sun}, C.~{She}, C.~{Yang}, T.~Q.~S. {Quek}, Y.~{Li}, and B.~{Vucetic},
  ``Optimizing resource allocation in the short blocklength regime for
  ultra-reliable and low-latency communications,'' \emph{IEEE Trans. Wireless
  Commun.}, vol.~18, no.~1, pp. 402--415, Jan 2019.

\bibitem{8756746}
W.~R. {Ghanem}, V.~{Jamali}, Y.~{Sun}, and R.~{Schober}, ``Resource allocation
  for multi-user downlink {URLLC-OFDMA} systems,'' in \emph{Proc. IEEE ICC
  Workshops}, May 2019, pp. 1--6.

\bibitem{3gppTR21915}
{3GPP}, ``{T}echnical specification group services and system aspects,'' {TR
  21.915 V15.0.0 }, Tech. Rep., September 2019.

\bibitem{6156468}
C.~{Shen}, T.~{Chang}, K.~{Wang}, Z.~{Qiu}, and C.~{Chi}, ``Distributed robust
  multicell coordinated beamforming with imperfect {CSI}: An {ADMM} approach,''
  \emph{IEEE Trans. Signal Process.}, vol.~60, no.~6, pp. 2988--3003, June
  2012.

\bibitem{8624559}
J.~{Chen}, L.~{Zhang}, Y.~{Liang}, X.~{Kang}, and R.~{Zhang}, ``Resource
  allocation for wireless-powered {IoT} networks with short packet
  communication,'' \emph{IEEE Trans. Wireless Commun.}, vol.~18, no.~2, pp.
  1447--1461, Feb 2019.

\bibitem{8345745}
X.~{Sun}, S.~{Yan}, N.~{Yang}, Z.~{Ding}, C.~{Shen}, and Z.~{Zhong},
  ``Short-packet downlink transmission with non-orthogonal multiple access,''
  \emph{IEEE Trans. Wireless Commun.}, vol.~17, no.~7, pp. 4550--4564, July
  2018.

\bibitem{8909370}
Y.~{Xu}, C.~{Shen}, T.~{Chang}, S.~{Lin}, Y.~{Zhao}, and G.~{Zhu},
  ``Transmission energy minimization for heterogeneous low-latency {NOMA}
  downlink,'' \emph{IEEE Trans. Wireless Commun.}, pp. 1--1, 2019.

\bibitem{5452208}
Y.~{Polyanskiy}, H.~V. {Poor}, and S.~{Verdu}, ``Channel coding rate in the
  finite blocklength regime,'' \emph{IEEE Trans. Inf. Theory}, vol.~56, no.~5,
  pp. 2307--2359, May 2010.

\bibitem{6802432}
W.~{Yang}, G.~{Durisi}, T.~{Koch}, and Y.~{Polyanskiy}, ``Quasi-static
  multiple-antenna fading channels at finite blocklength,'' \emph{IEEE Trans.
  Inf. Theory}, vol.~60, no.~7, pp. 4232--4265, July 2014.

\bibitem{8070468}
C.~{She}, C.~{Yang}, and T.~Q.~S. {Quek}, ``Cross-layer optimization for
  ultra-reliable and low-latency radio access networks,'' \emph{IEEE Trans.
  Wireless Commun.}, vol.~17, no.~1, pp. 127--141, Jan 2018.

\bibitem{7463506}
S.~{Xu}, T.~{Chang}, S.~{Lin}, C.~{Shen}, and G.~{Zhu}, ``Energy-efficient
  packet scheduling with finite blocklength codes: Convexity analysis and
  efficient algorithms,'' \emph{IEEE Trans. Wireless Commun.}, vol.~15, no.~8,
  pp. 5527--5540, Aug 2016.

\bibitem{8269136}
Y.~{Xu}, C.~{Shen}, T.~{Chang}, S.~{Lin}, Y.~{Zhao}, and G.~{Zhu},
  ``Energy-efficient non-orthogonal transmission under reliability and finite
  blocklength constraints,'' in \emph{Proc. IEEE Globecom Workshops}, Dec 2017,
  pp. 1--6.

\bibitem{Cand2007Enhancing}
E.~J. Candès, M.~B. Wakin, and S.~P. Boyd, ``Enhancing sparsity by reweighted
  $\ell_1$ minimization,'' \emph{Journal of Fourier Analysis \& Applications},
  vol.~14, no.~5, pp. 877--905, 2007.

\bibitem{6365845}
W.~{Li}, T.~{Chang}, C.~{Lin}, and C.~{Chi}, ``Coordinated beamforming for
  multiuser {MISO} interference channel under rate outage constraints,''
  \emph{IEEE Trans. Signal Process.}, vol.~61, no.~5, pp. 1087--1103, March
  2013.

\bibitem{Razaviyayn2012A}
M.~Razaviyayn, M.~Hong, and Z.~Q. Luo, ``A unified convergence analysis of
  block successive minimization methods for nonsmooth optimization,''
  \emph{SIAM J. Optim.}, vol.~23, no.~2, pp. 1126--1153, 2012.

\end{thebibliography}


{\small

}

\end{document}